\documentclass[preprint,1p,groupedaddress]{elsarticle}
\usepackage{graphicx}  
\usepackage{dcolumn}   
\usepackage{bm}        
\usepackage{amssymb}   
\usepackage{slashed}
\usepackage{graphicx}				
\usepackage{amsmath}
\usepackage{tikz}
\usetikzlibrary{arrows,backgrounds}
\usepackage[all]{xy}

\begin{document}
\title{A field theoretical approach to the P vs. NP problem via the phase sign of Quantum Monte Carlo}
\author[rvt]{Andrei T. Patrascu}

\address[rvt]{University College London, Department of Physics and Astronomy, London, WC1E 6BT, UK}
\begin{abstract}
I present here a new method that allows the introduction of a discrete auxiliary symmetry in a theory in such a way that the eigenvalue spectrum 
of the fermion functional determinant is made up of complex conjugated pairs. The method implies a particular way of introducing and integrating
 over auxiliary fields related to a set of artificial shift symmetries.  Gauge-fixing the artificial continuous shift symmetries in the direct and dual sectors leads to the implementation of direct and dual BRST-type global symmetries and of a symplectic structure over the field space (as prescribed by the Batalin-Vilkovisky method).
 A procedure similar to Kahler polarization in geometric quantization guarantees the possibility to choose a Kahler structure over the field space.
This structure is generated by a special way of performing gauge fixing over the direct and dual sectors.
 The desired discrete symmetry appears to be induced by the Hodge-* operator. The particular extension of the field space presented
 here makes the operators of the de-Rham cohomology manifest. These become symmetries in the extended theory.
This method implies the identification of the (anti)-BRST and dual-(anti)-BRST operators with the exterior derivative and its dual in the context
 of the complex de-Rham cohomology. The novelty of this method relies on the fact that the field structure is doubled two times in order to make
 use of a supplemental symmetry prescribed by algebraic geometry. Indeed every auxiliary field configuration presents this symmetry as the auxiliary fields themselves are related via the dual sector of the BRST-(anti)-BRST transformations. 
 This leads to a generalization of Kramers theorem that avoids the Quantum Monte Carlo phase sign problem without any apparent increase in complexity. The applicability of this method in the case of strongly coupled fermions makes it a possible example of the use of strong-weak dualities for field spaces. 
\end{abstract}
\address{andrei.patrascu.11@ucl.ac.uk}
\maketitle
\section{Introduction}
The $P$ vs. $NP$ problem is known to have significant implications in many areas of science, not excluding physics, mathematics or information theory [1]. 
The main question is if some specific classes of problems can be efficiently treated via algorithmic methods. In this paper I present a constructive 
approach to this problem based essentially on some topological and geometrical arguments. While researchers speculate about this question [2] and give several interpretations 
of possible results there have been very few attempts to analyze the problem from perspectives other than purely algorithmic.
As shown in ref. [3] the Quantum Monte Carlo phase sign problem can be mapped into a general $NP$-complete problem. I will follow this paper for
 a short introduction into the statistical aspects of the subject.
The idea behind Monte Carlo simulations is to replace the direct calculation of sums of the form
\begin{equation}
\begin{array}{ll}
<A>=\frac{1}{Z}\sum_{c\in\Omega}A(c)p(c); & Z=\sum_{c\in\Omega}p(c)\\
\end{array}
\end{equation}
over a high dimensional space $\Omega$ of configurations $\{c\}$ with the sum over a set of $M$ configurations $\{c_{i}\}$ from $\Omega$ according
 to the distribution $p(c_{i})$.
The average is then calculated as 
\begin{equation}
<A>\approx\bar{A}=\frac{1}{M}\sum_{i=1}^{M}A(c_{i})
\end{equation}
The statistical error of the above calculation is given by 
\begin{equation}
\Delta A=\sqrt{Var_{A}(2\tau_{A}+1)/M}
\end{equation}
$Var_{A}$ being the variance of A and $\tau_{A}$ measures the autocorrelations of the sequence $\{A(c_{i})\}$.

The Monte Carlo approach permits the evaluation of the same average in polynomial time as long as $\tau_{A}$ does not increase faster than polynomial in the number of 
particles. 
For physical systems the sum one needs to calculate changes as follows
\begin{equation}
\begin{array}{ll}
<A>=\frac{1}{Z}Tr[A exp(-\beta H)]; & Z=Tr(exp(-\beta H))
\end{array}
\end{equation}
where $\beta$ is the inverse temperature and $Z$ is the partition function.
A Monte Carlo technique can still be applied to reduce the exponential scaling of the problem, but, as specified in [3], only after the mapping of the quantum model 
on a classical one. The nature of this mapping is considered by [3], following [4], to be 
a Taylor expansion. 
\begin{equation}
Z=Tr (exp(-\beta H))=\sum_{n=0}^{\infty}\frac{{-\beta}^{n}}{n!}Tr(H^{n})=
\end{equation}
\begin{equation}
=\sum_{n=0}^{\infty}\sum_{i_{1},...i_{n}}\frac{{-\beta}^{n}}{n!}<i_{1}|H|i_{2}>...<i_{n}|H|i_{1}>=
\end{equation}
\begin{equation}
=\sum_{n=0}^{\infty}\sum_{i_{1},...i_{n}}p(i_{1},...i_{n})=\sum_{c}p(c)
\end{equation}
For each order $n$ in the expansion, $n$ sums were inserted over a complete basis set of states $\{|i>\}$.
The configurations are sequences of $n$ basis states and the weight $p(c)$ is associated to the summand above. 
The average becomes now
\begin{equation}
<A>=\frac{1}{Z}Tr[A exp(-\beta H)]=\frac{1}{Z}\sum_{c}A(c)p(c)
\end{equation}
As long as the weight $p$ is positive a standard Monte Carlo technique can be applied. In fermionic systems this is not true as negative weights are possible.
It is argued in [3] that although a change of the basis $\{|i>\}$ that makes the weights always positive is possible, the complexity of the method needed to 
find the required
 transformation must be exponential. Also, the authors of [3] map the sign problem into a problem that is $NP$-complete. This is of course correct if one follows the above
 steps. The main scope of this paper is to prove that some assumptions in [3]
can be avoided when considering a different quantization prescription
and that a system can be mapped into an $NP$-complete problem but still have a polynomial solution if analyzed from the perspective of 
the quantization of gauge theories. A gauge symmetry can be seen as a redundancy of the mathematical formulation. As shown by Batalin and Vilkovisky in [5] if one is willing to lose the explicit visibility of some properties one can reduce the gauge symmetry and transform a gauge theory into a non-gauge one. In this paper, I will follow the opposite path.
 I construct a theory that has artificial gauge symmetries introduced in such a way that a discrete symmetry to be associated with an artificial "time reversal" invariance appears. Following ref.
 [6] the presence
 of such a symmetry in a theory permits the avoidance of the sign problem. This construction is done by using the field-antifield [7] quantization of gauge theories with general algebras.

It can be argued that the time reversal type symmetry is not visible when looking at each configuration separately. However, from the construction, the individual fields are related via symmetry transformations with the additional fields in the "bulk field space". Hence, as will be seen further on, for each original configuration, there exists a bulk field configuration which obeys the required symmetry rules while being in the same BRST-anti-BRST-co-(anti)-BRST class. One may think that this method can be applied only in situations when the fermionic fields are related by a time reversal type symmetry originally and that these fields are usually strongly coupled. However, there exists a duality: strongly coupled fermionic fields (with no obvious symmetry) can be mapped into fermionic fields related via time reversal type symmetry with fictitious fields. In this way, even after decoupling (in the strongly coupled sector), the symmetry can be implemented into the fictitious field sector maintaining the same physics inside the problem. A strongly coupled problem in the sector of fields commonly used, can be associated with a "weakly coupled" problem in the "bulk field space" with additional fictitious supplemental fields and symmetries. The original theory with strongly coupled fermions becomes a theory with a symmetry between the original fermionic fields and the fictitious "bulk field space" components. In this sense, this appears to be an application of strong-weak dualities in the case of quantum monte carlo calculations. Hence, the symmetry becomes a dynamical object, behaving in such a way that it mediates what was the strong coupling in the original theory via a special configuration of bulk fields. 

In order to continue, I partially follow the description by Alfaro and Damgaard [8],[9] in order to show what is
 the effect of the 
quantization of a field theory with fermions, how the change in sign appears and how one can relate classical and quantum descriptions in a different way. I also make the
 connection between 
geometry (symmetry) and topology (cohomology) by introducing the BRST, anti-BRST and dual-(anti)BRST [10] operators associated to the de-Rham cohomology.  I define and use the 
Hodge star operation [11]
 in this context in order to generate a discrete symmetry. I make use of the intrinsic symplectic structure of the general field-antifield functional space in order
to generate a Kahler structure([12],[13]). I also use the fact
 that the
 extension of the field space towards an even dimensional space is always possible. The end result is a general quantum field theory free of the Monte Carlo sign problem
 and with no 
apparent exponential growth in complexity.
\section{Quantization Prescriptions}
The idea of quantization has a vast history. Originally, physical variables have been promoted to operators with specific commutation rules.
 These encoded the first quantizations ever performed. They were followed by the second quantization prescriptions and the anti-commutators needed for the description
 of fermionic particles. Finally path integral quantization brought a completely new perspective on the procedure of quantization.
While a classical theory is described by an action functional and a minimization prescription, a path integral quantization is constructed as a functional integral 
of the complex exponentiated action functional
\begin{equation}
 exp(iS[.]):C\rightarrow A
\end{equation}
where $C$ is the configuration space and $A$ is the resulting space. This definition is very formal. In practical situations the measure of the path integral
is not always defined in the standard way. The configuration spaces are in general not even manifolds. Sometimes, in order to obtain pertinent results a so
called ``cohomological integration'' is necessary.
 When the theory we want to
quantize has redundancies (gauge symmetries) one relies
on two possible approaches. When the gauge algebra
is closed a BRST quantization procedure can be implemented. In general however, the gauge algebra does not
close. In this case an alternative method developed initially by Batalin and Vilkovisky is used. 

The algebra of the operators of the gauge symmetry can in general be defined as

\begin{equation}
 \frac{\delta^{l}R^{i}_{\alpha}}{\delta\phi^{j}}R_{\beta}^{j}-(-1)^{\epsilon_{\alpha}\epsilon_{\beta}}\frac{\delta^{l}R^{i}_{\beta}}{\delta\phi^{j}}R_{\alpha}^{j}=2R_{\gamma}^{i}T_{\alpha\beta}^{\gamma}(-1)^{\epsilon_{\alpha}}-4y_{j}E^{ji}_{\alpha\beta}(-1)^{\epsilon_{i}}(-1)^{\epsilon_{\alpha}}
\end{equation}

where $y_{j}=0$ represents the equation of motion, $E$ and $T$ represent coefficients, $R$ represent the (gauge) symmetry transformation operators and $\epsilon$ encodes the 
Grassmann parity of the associated field.
One can also define the BRST transformations of the original fields as $\delta\phi^{i}=R_{\alpha}^{i}[\phi]c^{\alpha}$ i.e. one can define the BRST symmetry transformations
via $R[\phi]$ and the associated ghost field $c^{\alpha}$ unambiguously. This is why, when no confusion is possible the terms 
$R^{i}_{\alpha}$, $R[\phi^{i},c,...]$ or the BRST transformation rule $\delta\phi^{A}=R^{A}[\phi^{B}]$ will be used alternatively as formal definitions. 

If $E=0$ the algebra is closed and the nilpotency of the BRST operator is naively verified. Imposing nilpotency on the fields $\phi^{i}$ we get
\begin{equation}
  0=\delta^{2}\phi^{i}=R_{\alpha}^{i}\delta c^{\alpha}+\frac{\delta^{l}R_{\alpha}^{i}c^{\alpha}}{\delta\phi^{j}}R_{\beta}^{j}c^{\beta}\\
\end{equation}
If we choose now 
\begin{equation}
 \delta c^{\gamma}=T^{\gamma}_{\alpha\beta}[\phi]c^{\beta}c^{\alpha}
\end{equation}
the nilpotency condition on the ``physical'' sector is satisfied and we obtain (considering $E=0$) 
\begin{equation}
 \frac{\delta^{l}R_{\alpha}^{i}c^{\alpha}}{\delta\phi^{j}}R_{\beta}^{j}c^{\beta}+R_{\gamma}^{i}T^{\gamma}_{\alpha\beta}c^{\beta}c^{\alpha}=0
\end{equation}
Also, using Jacobi identity one can easily show that $\delta^{2}c^{\gamma}=0$.
It will be seen later how this can be generalized for the case of BRST-anti-BRST transformations. 
If the algebra depends on the last term i.e. $E$ is not zero we have an open algebra and an non-nilpotent BRST transformation as acting on the initial fields.
The gauge fixed action constructed in the naive way would not be BRST invariant off-shell.
In order to solve this problem one had to introduce an artificial shift symmetry and to move the non-nilpotency from the transformation rules of the original fields
to the transformation rules of the collective fields. 
One certainly trivial way of enlarging the field space is by introducing two fields $A^{l}$ and $B^{l}$ such that
\begin{equation}
\begin{array}{l} 
\delta A^{l}=B^{l}\\
\delta B^{l}=0\\
\end{array}
\end{equation}
Obviously as the initial action does not depend on $A^{l}$ one can shift it with no practical effect. This shift would be a local symmetry and the fields $B^{l}$ 
would be the associated ghost-fields. It is precisely this idea that allows the redefinition of the 
field structure as will be seen further on. 
While it is certainly possible to move undesirable aspects of the theory to the collective sector it is also possible to transfer desirable properties to the field structure 
while keeping the well behaved properties inside. 
Moreover, if there are more symmetries then the interplay between them at the level of the BRST (-anti-BRST-dual-(anti)-BRST) introduces additional freedoms that I am 
using in order to avoid the sign problem practically permanently.

As one can see
by now, the quantization prescription is not always trivial. One must specify what quantization means in the
framework of path integrals. Essentially the special way
in which the functional integration is performed assures
the correct quantization of a classical theory. Moreover,
the theory, defined by an action functional is by no means
unique. It is well known that different representations
can be chosen but in general in physics this amounts to
the construction of effective low energy theories. This
doesn’t always have to happen in this way. The chosen
field structure can be designed such that it maps a complexity class into another.

I start with a general field theory as described by the action $S[\phi^{A}]$. The Batalin-Vilkovisky quantization prescription enlarges the field-space of the theory
 by introducing 
antifields ($\phi_{A}^{*}$) and gives a new canonical structure known as the antibracket [14].
This is defined considering two Grassmann functionals $F$ and $G$ as 

\begin{equation}
(F,G)=\frac{\delta^{r}F}{\delta\phi^{A}(x)}\frac{\delta^{l}G}{\delta\phi^{*}_{A}(x)}-\frac{\delta^{r}F}{\delta\phi^{*}_{A}(x)}\frac{\delta^{l}G}{\delta\phi^{A}(x)}
\end{equation}
involving alternate functional differentiation with respect to the fields and antifields. $r$ and $l$ superscripts stand for the right and left derivatives respectively.
I am following here the rules of reference [8] for the left and right derivatives. Accordingly
\begin{equation}
 \frac{\delta^{l}(FG)}{\delta{A}}=\frac{\delta^{l}F}{\delta{A}}G+(-1)^{\epsilon_{F}\epsilon_{A}}F\frac{\delta^{l}G}{\delta{A}}
\end{equation}
\begin{equation}
 \frac{\delta^{r}(FG)}{\delta{A}}=F\frac{\delta^{r}G}{\delta{A}}+(-1)^{\epsilon_{G}\epsilon_{A}}\frac{\delta^{r}F}{\delta{A}}G
\end{equation}
which amount to the following relation between left and right derivatives in general
\begin{equation}
 \frac{\delta^{l}F}{\delta{A}}=(-1)^{\epsilon_{A}(\epsilon_{F}+1)}\frac{\delta^{r}F}{\delta{A}}
\end{equation}
The antibracket has some important properties: it changes the statistics as 
\begin{equation}
\epsilon[(F,G)]=\epsilon(F)+\epsilon(G)+1
\end{equation}
and satisfies the following relation
\begin{equation}
(F,G)=-(-1)^{(\epsilon(F)+1)(\epsilon(G)+1)}(G,F)
\end{equation}
where $\epsilon$ is the Grassmann parity operator.
Using this structure the Batalin-Vilkovisky prescription can be written as
\begin{equation}
\frac{1}{2}(W,W)=i\hbar\Delta W
\end{equation}
where 
\begin{equation}
\Delta=(-1)^{\epsilon_{A}+1}\frac{\delta^{r}}{\delta\phi^{A}}\frac{\delta^{r}}{\delta\phi^{*}_{A}}
\end{equation}
W is called the "quantum action" and is a solution of the above equation. If it can be expanded in powers of $\hbar$ one obtains:
\begin{equation}
W=S+\sum_{n=0}^{\infty}\hbar^{n}M_{n}
\end{equation}
The boundary conditions should make this coincide with the classical action when all antifields are removed $(\phi_{A}^{*}=0)$. 
To the lowest order one recovers the classical master equation $(S,S)=0$.

If one starts with the classical action (containing the usual number of fields)  $S[\phi^{A}]$ the associated path integral is 
\begin{equation}
Z=\int [d\phi^{A}]exp[\frac{i}{\hbar}S[\phi^{A}]]
\end{equation}
By performing the transformations $\phi^{A}(x)\rightarrow\phi^{A}(x)-\varphi^{A}(x)$ one constructs an action 
 $S[\phi^{A}-\varphi^{A}]$ invariant to a local shift symmetry
\begin{equation}
\begin{array}{l}
\delta\phi^{A}(x)=\Theta(x)\\
\delta\varphi^{A}(x)=\Theta(x)\\
\end{array}
\end{equation}
where $\Theta(x)$ is arbitrary. 
In this way I constructed another field representation that contains a collective field $\varphi^{A}$. One can in principle integrate over the collective field if one 
fixes the introduced gauge symmetry
 in the standard BRST manner: add an BRST-exact term in such a way that the local gauge symmetry is broken. This term must contain a 
ghost-antighost pair $(c^{A}(x),\phi_{A}^{*}(x))$ and a
 Nakanishi-Lautrup field $B_{A}(x)$. A global BRST symmetry should emerge.
The transformation rules of the fields in this theory will be
\begin{equation}
\begin{array}{l}
\delta\phi^{A}(x)=c^{A}(x) \\
\delta\varphi^{A}(x)=c^{A}(x) \\
\delta c^{A}(x)=0 \\
\delta\phi_{A}^{*}(x)=B_{A}(x) \\
\delta B_{A}(x)=0 \\
\\
\\
\end{array}
\end{equation}
I make no assumptions about the Grassmann parity of the initial fields $\phi^{A}$.
The ghost numbers of the new fields will be
\begin{equation}
\begin{array}{lll}
gh(c^{A})=1; & gh(\phi_{A}^{*})=-1; & gh(B_{A})=0\\
\end{array}
\end{equation}
and $\delta$ is statistics changing. 
One can gauge fix the transformed action by adding
\begin{equation}
-\delta[\phi_{A}^{*}\varphi^{A}]=(-1)^{\epsilon(A)+1}B_{A}\varphi^{A}-\phi_{A}^{*}c^{A}
\end{equation}
where $\epsilon(A)$ is the Grassmann parity of the field $\phi^{A}$. 
The partition function is now well defined 

\begin{equation}
\begin{array}{l}
Z=\int[d\phi_{A}][d\varphi_{A}][d\phi^{*}_{A}][dc_{A}][dB_{A}]\\
exp[\frac{i}{\hbar}(S[\phi_{A}-\varphi_{A}]-\int dx[(-1)^{\epsilon(A)}B_{A}(x)\varphi^{A}(x)+\phi^{*}_{A}(x)c^{A}(x)])]\\
\end{array}
\end{equation}

The collective field has been gauge fixed to zero. If one integrates out $B_{A}(x)$ one obtains
\begin{equation}
\begin{array}{l}
Z=\int[d\phi^{A}][d\phi_{A}^{*}][dc^{A}]exp(\frac{i}{\hbar}S_{ext})\\
S_{ext}=S[\phi^{A}]-\int[dx]\phi^{*}_{A}c^{A}(x)
\end{array}
\end{equation}
where the ghosts are decoupled.
From here $\frac{\delta^{r}S_{ext}}{\delta\phi_{A}^{*}}=-c^{A}(x)$ or similarly $\frac{\delta^{l}S_{ext}}{\delta\phi_{A}^{*}}=c^{A}(x)$.
Substituting now the field equation of motion for $B_{A}(x)$ we obtain the symmetry transformations
\begin{equation}
\begin{array}{l}
\delta\phi^{A}(x)=c^{A}(x)\\
\delta c^{A}(x)=0\\
\delta \phi_{A}^{*}(x)=-\frac{\delta^{l}S}{\delta \phi^{A}(x)}\\
\end{array}
\end{equation}
where the superscripts $l$ and $r$ represent the left and right derivatives respectively.
This symmetry generates the Schwinger Dyson equations. Starting from the identity $0=<\delta\{\phi_{A}^{*}(x)F[\phi^{A}]\}>$ and integrating over the ghosts $c^{A}$ and the antighosts $\phi_{A}^{*}$ the Ward identity becomes
\begin{equation}
<\frac{\delta^{l}F}{\delta\phi^{A}(x)}+(\frac{i}{\hbar})\frac{\delta^{l}S}{\delta\phi^{A}(x)}F[\phi^{A}]>=0
\end{equation}
which is the most general Schwinger-Dyson equation to be associated to this theory. 
Now, the equation that expresses the BRST invariance of the extended action is
\begin{equation}
\begin{array}{l}
0=\delta S_{ext}=\int dx \frac{\delta^{r}S_{ext}}{\delta\phi^{A}(x)}c^{A}(x)-\int dx \frac{\delta^{r}S_{ext}}{\delta\phi^{*}_{A}(x)}\frac{\delta^{l}S}{\delta\phi^{A}(x)}=\\
=\int dx \frac{\delta^{r}S_{ext}}{\delta\phi^{A}(x)}c^{A}(x)-\int dx \frac{\delta^{r}S_{ext}}{\delta\phi^{*}_{A}(x)}\frac{\delta^{l}S_{ext}}{\delta\phi^{A}(x)}\\
\end{array}
\end{equation}
where $S$ differs from $S_{ext}$ by a term independent of $\phi^{A}$.

Using the definition of the antibracket written in general for two functionals $F$ and $G$ as
\begin{equation}
(F,G)=\frac{\delta^{r}F}{\delta\phi^{A}(x)}\frac{\delta^{l}G}{\delta\phi^{*}_{A}(x)}-\frac{\delta^{r}F}{\delta\phi^{*}_{A}(x)}\frac{\delta^{l}G}{\delta\phi^{A}(x)}
\end{equation}
 the above identity corresponds to what is called the master equation 
\begin{equation}
\frac{1}{2}(S_{ext},S_{ext})=-\int dx\frac{\delta^{r}S_{ext}}{\delta\phi^{A}(x)}c^{A}(x)
\end{equation}
The important aspect to be considered here is the right hand side term of the above formula. In this case the solution of the above expression implies an expansion in terms of the ghosts and the antighosts with a set of unknown coefficients
\begin{equation}
S_{ext}[\phi^{A},\phi_{A}^{*},c^{A}]=S[\phi^{A}]+\sum_{n=1}^{\infty}a_{n}\phi_{A_{1}}^{*}...\phi_{A_{n}}^{*}c^{A_{1}}...c^{A_{n}}
\end{equation}
Of course the choice of integrating over both the ghosts and the anti-ghosts is arbitrary. One can chose to integrate only over the ghost fields $c^{A}(x)$ but not over the corresponding anti-ghosts $\phi_{A}^{*}(x)$. The partition function becomes then
\begin{equation}
Z=\int[d\phi^{A}][d\phi_{A}^{*}]\delta(\phi_{A}^{*})exp[\frac{i}{\hbar}S[\phi^{A}]]
\end{equation}
On the side of the BRST algebra this change amounts in the way in which the non-propagating fields are replaced by their corresponding equations of motion. 
The direct method used above must be refined when dealing with fermionic type fields. 
The main question is how to replace $c^{A}$ inside the Green functions? The answer to this question will give the transformation rules for the fields that are not integrated out. 
Consider the identity 

\begin{equation}
\begin{array}{l}
\int[dc]F[c^{B}(y)]exp[-\frac{i}{\hbar}\int dx\phi_{A}^{*}(x)c^{A}(x)]=\\
F(i\hbar\frac{\delta^{l}}{\delta\phi_{B}^{*}(y)})exp[-\frac{i}{\hbar}\int dx \phi_{A}^{*}(x)c^{A}(x)]\\
\end{array}
\end{equation}

It follows that the replacement of $c^{A}$ with its equation of motion $(c^{A}(x)=0)$ is not sufficient. One has to add what is called a "Quantum Correction" of the form
$\hbar\delta/\delta\phi^{*}$. What appears as "Quantum Correction" in the Green function results from our choice of integrating only over a ghost field and not 
over its associated anti-ghost. Essentially it is at this point in the quantization procedure where the difference between fermionic and bosonic fields appears.
The BRST symmetry transformations have to change accordingly if the option of integrating only over the fermionic "half" of the field-antifield structure is chosen. 
Now, performing this replacement (meaningful only inside the path integral) the BRST transformation itself becomes
\begin{equation}
\begin{array}{l}
\delta\phi^{A}(x)=i\hbar(-1)^{\epsilon_{A}}\frac{\delta^{r}}{\delta\phi_{A}^{*}(x)}\\
\delta\phi^{*}_{A}(x)=-\frac{\delta^{l}S}{\delta\phi^{A}(x)}\\
\end{array}
\end{equation}
where $\epsilon_{A}$ is the Grassmann parity associated to the fields indexed by $A$.
It can be checked that this transformation leaves at least the combination of the measure and the action invariant. 
After integrating out the ghost the antibracket structure is modified. In order to see how, one can perform a variation of an arbitrary
 functional $G[\phi^{A},\phi_{A}^{*}]$. Inside the path integral we obtain

\begin{equation}
\begin{array}{l}
\delta G[\phi^{A},\phi_{A}^{*}]\\
=\int dx \frac{\delta^{r}G}{\delta\phi^{A}(x)}[\frac{\delta^{l}S_{ext}}{\delta\phi_{A}^{*}(x)}+(i\hbar)(-1)^{\epsilon_{A}}\frac{\delta^{r}}{\delta\phi^{*}_{A}(x)}]-\int dx \frac{\delta^{r}G}{\delta\phi^{*}_{A}(x)}\frac{\delta^{l}S_{ext}}{\delta\phi^{A}(x)}\\
\end{array}
\end{equation}

The left derivative on the left side comes from the definition of $c^{A}$ in $S_{ext}$. In this case it can be considered simply as zero. I follow here closely the notation of
reference [8]. This equation describes the "quantum deformation" introduced in the antibracket structure
\begin{equation}
\delta G[\phi^{A},\phi_{A}^{*}]=\int dx [(G,S_{ext})-i\hbar\Delta G]
\end{equation}
where
\begin{equation}
\Delta=(-1)^{\epsilon_{A}+1}\frac{\delta^{r}\delta^{r}}{\delta \phi^{*}_{A}(x)\delta\phi^{A}(x)}
\end{equation}
Again, this term appears only as a consequence of the partial integration which allows us to expose the operator $\delta / \delta\phi_{A}^{*}$ that otherwise acts only on a $\delta$-functional. Whenever one choses to keep only half of the field-antifield components in the theory and integrates over the ghosts obeying a fermi statistics the result will be a deformation of the antibracket structure that will lead to the Quantum Master Equation. 
One can already see that the mapping of the "quantum" problem to the "classical" problem as presented in reference [3] is correct but not unique. 
In the next chapter I show how one should extend the field structure of an arbitrary theory in order to obtain a sign-problem free theory. This prescription implies by no means any exponential increase in complexity if one considers what has been presented above.

\section{Construction of the theory}
\par My method relies at a first level on the field-antifield formalism as introduced by Batalin and Vilkovisky [5] and at a second level on an innovative use of some algebraic geometry and topology theorems [11].
I will regard the partition function as depending on the action functional described in terms of a set of fields $\phi^{A}$
\begin{equation}
Z=Z_{0}\int exp(-\sum_{i}S_{i}[\phi^{A}]) D\phi^{A}
\end{equation}
More generally, the theory may have additional internal symmetries, generated by corresponding operators. 
First, let me show here the main idea related to the introduction of a single shift symmetry using one set of collective fields. 
I will regard the partition function as depending on the action functional described in terms of a set of fields $\phi^{A}$ (43). 
More generally, the theory may have additional internal symmetries, generated by corresponding operators. 
Let me now double the fields by introducing a collective field $\varphi^{A}$ that induces a shift symmetry in the theory:
\begin{equation}
 \phi^{A}\rightarrow\phi^{A}-\varphi^{A}
\end{equation}
No assumption regarding the statistics of the $\phi^{A}$ fields is required. They can be fermionic or bosonic. The new shift symmetry must be gauge fixed and for
 this I have to introduce a ghost and a trivial system in the form of a multiplet consisting of an antifield $\phi^{*}_{A}$ and an auxiliary field $B_{A}$.
 After gauge fixing a global BRST symmetry emerges in general. In the present case however, in order to maintain the triviality of the extended BRST symmetry 
that encompasses also the shift symmetry the BRST transformation rules change in the following way:
\begin{equation}
\begin{array}{l}
 \delta\phi^{A}=c^{A}\\
 \delta\varphi^{A}=c^{A}-R^{A}[\phi^{A}-\varphi^{A}]\\
 \delta c^{A}=0\\
 \delta\phi^{*}_{A}=B_{A}\\
 \delta B_{A}=0\\
\end{array}
\end{equation}
where $c^{A}$ is the ghost field and $\delta$ is the BRST transformation to be associated with the total BRST-type symmetries. 
$R^{A}[\phi^{A}-\varphi^{A}]$ represents a formal definition of the BRST symmetry associated to the possible intrinsic initial gauge symmetry.
There exist a freedom to shift the original gauge symmetry $R^{A}[\phi^{A}-\varphi^{A}]$ between the BRST transformation of the fields and the BRST transformation
of the collective fields. In this way a possible off-shell non-nilpotency in the transformation rules is transfered to the transformation rules of the collective field.
This was one of the first applications of the collective field formalism to the quantization of theories involving algebras that do not close (i.e. algebras of the
 gauge symmetry generators depending on the form of the field equations of motion). As an example of such theories one may quote supergravity.
The ghost numbers of the new fields are
\begin{equation}
\begin{array}{lll}
gh(c^{A})=1; & gh(\phi_{A}^{*})=-1; & gh(B_{A})=0
\end{array}
\end{equation}
Although the new artificial and certainly trivial 
continuous shift symmetry is easy to eliminate at this level, it is of major importance as a tool for generating new discrete symmetries. 
Considering the new continuous shift symmetry, one has to gauge fix it. This can be done adding the following terms in form of BRST transformations:
\begin{equation}
\begin{array}{l}
 S_{gf}=S_{0}[\phi^{A}-\varphi^{A}]-\delta[\phi^{*}_{A}\phi^{A}]+\delta \Psi[\phi^{A}]=\\
 =S_{0}[\phi^{A}-\varphi^{A}]+\phi_{A}^{*}R^{A}[\phi^{A}-\varphi^{A}]-\phi_{A}^{*}c^{A}+\frac{\delta^{l}\Psi}{\delta\phi^{A}}c^{A}-\varphi^{A}B_{A}=\\
 =S_{BV}[\phi^{A}-\varphi^{A}]-\phi_{A}^{*}c^{A}+\frac{\delta^{l}\Psi}{\delta\phi^{A}}c^{A}-\varphi^{A}B_{A}\\
\end{array}
\end{equation}
where $S_{BV}$ is called the Batalin-Vilkovisky action. It incorporates the original action and the terms arising from other possible internal gauge symmetries.
$\Psi$ is a gauge fixing bosonic functional depending only on the original fields.  
By this I define a new gauge fixed action. Note that the nillpotency $\delta^{2}=0$ of the BRST transformation assures the overall invariance. 
The partition function is the standard one:
\begin{equation}
 Z=\int[d\phi^{A}][d\phi^{*}_{A}]\delta(\phi^{*}_{A}-\frac{\delta^{l}\Psi[\phi^{A}]}{\delta\phi^{A}})e^{-S_{BV}[\phi^{A},\phi^{*}_{A}]}
\end{equation}
Where $\Psi[\phi^{A}]$ is defined considering the condition imposed in the resulting delta-function.
One must underline that the gauge fixing procedure must keep the gauge independence of the full partition function including the integration measure. 
\par Until now, a continuous shift-symmetry has been introduced and gauge fixed.

However, the action presents further flexibility. One can extend the field structure such that two BRST operators become manifest. In this way one implements the
 BRST-anti-BRST symmetry and the associated field structure [7]. This method allows Schwinger-Dyson equations as Ward identities as well.
Moreover, this method also known as the ``$Sp(2)''$-invariant quantization has the property of manifestly generating a symplectic structure over the field space. 
This will prove to be important further on. 
 The method is similar to what has been shown before and we obtain
$S_{0}[\phi]\rightarrow S_{0}[\phi_{A}-\varphi_{A1}-\varphi_{A2}]$.
Two extra gauge symmetries arise for which a new structure of fields is introduced: two ghostfields ($c_{A1},\phi^{*}_{A2}$) and two antighost fields ($\phi^{*}_{1},c_{A2}$).
Of course this extends the symmetries allowed in the theory.
\begin{equation}
\begin{array}{llll}
   \delta_{1}\phi_{A}=c_{A1} & &  & \delta_{2}\phi_{A}=c_{A2} \\
   \delta_{1}\varphi_{A1}=c_{A1}-\phi_{A2}^{*} & &  & \delta_{2}\varphi_{A1}=-\phi^{*}_{A1} \\
   \delta_{1}\varphi_{A2}=\phi^{*}_{A2} & &  & \delta_{2}\varphi_{A2}=c_{A2}+\phi^{*}_{A1} \\
   \delta_{1}c_{A1}=0 & &  & \delta_{2}c_{A2}=0\\
   \delta_{1}\phi_{A2}^{*}=0 & &  & \delta_{A2}\phi_{A1}^{*}=0\\
\end{array}
\end{equation}
Here $\delta_{1}$ and $\delta_{2}$ are respectively the BRST and anti-BRST transformations. 
The next step is to impose gauge fixing. This is done in the standard way by adding more bosonic fields, call them $B_{A}$ and $\lambda_{A}$.
The BRST transformation rules extend according to
\begin{equation}
 \begin{array}{llll}
  \delta_{1}c_{A2}=B_{A} & &  & \delta_{2}c_{A1}=-B_{A}\\
  \delta_{1}B_{A}=0 & &  & \delta_{2}B_{A}=0\\
  \delta_{1}\phi_{1}^{*}=\lambda_{A}-\frac{B_{A}}{2} & &  & \delta_{2}\phi_{2}^{*}=-\lambda_{A}-\frac{B_{A}}{2}\\
  \delta_{1}\lambda_{A}=0 & &  & \delta_{2}\lambda_{A}=0
 \end{array}
\end{equation}
These rules imply the nillpotency conditions $\delta_{1}^{2}=\delta_{2}^{2}=\delta_{1}\delta_{2}+\delta_{2}\delta_{1}=0$.
The action invariant under this BRST symmetry contains the terms of $S_{0}[\phi_{A}-\varphi_{A1}-\varphi_{A2}]$ plus some gauge fixing terms
\begin{equation}
\begin{array}{ll}
 S_{col}=\frac{1}{2}\delta_{1}\delta_{2}[\varphi_{A1}^{2}-\varphi_{A2}^{2}]=\\
 =-(\varphi_{A1}+\varphi_{A2})\lambda_{A}+\frac{B_{A}}{2}(\varphi_{A1}-\varphi_{A2})+(-1)^{a}\phi_{Aa}^{*}c_{Aa}\\
\end{array}
\end{equation}
Here summation over $a=1,2$ is implied. 
Using the transformation 
\begin{equation}
 \varphi_{A\pm}=\varphi_{A1}\pm\varphi_{A2}
\end{equation}
we obtain the gauge fixed action
\begin{equation}
 S_{gf}=S_{0}[\phi_{A}-\varphi_{A+}]-\varphi_{A+}\lambda_{A}+\frac{B_{A}}{2}\varphi_{A-}+(-1)^{a}\phi_{Aa}^{*}c_{Aa}
\end{equation}
For the sake of generality I follow the notation in reference [7] and define
\begin{equation}
 \delta_{a}\phi_{A}=R_{Aa}(\phi_{A})
\end{equation}
where $R_{Aa}$ is the BRST-anti-BRST symmetry transformation associated to an initial intrinsic gauge symmetry.
Apart from this, the transformations associated to the additional artificial shift symmetries will be added. 
In the case $a=1$ we have the BRST transformation rules whereas in the case $a=2$ we have the anti-BRST transformation rules.
The two collective fields are denoted by $\varphi_{A1}$ and $\varphi_{A2}$ or generally $\varphi_{Aa}$. The transformation will be $\phi_{A}-\varphi_{A1}-\varphi_{A2}$.
The field multiplets used are the ghosts $(c_{A1},\phi_{A}^{*2})$ and the antighosts $(\phi_{A}^{*1},c_{A2})$.
For $a=1$, $c_{Aa}$ is a ghost while for $a=2$, $c_{Aa}$ is an anthighost.
The BRST-anti-BRST transformations are 
\begin{equation}
\begin{array}{ll}
 \delta_{a}\phi_{A}=c_{Aa}\\
\delta_{a}\varphi_{Ab}=\delta_{ab}[c_{Aa}-\epsilon_{ac}\phi_{A}^{*c}-\\
 -R_{Aa}(\phi_{A}-\varphi_{A1}-\varphi_{A2})]+(1-\delta_{ab})\epsilon_{ac}\phi_{A}^{*c}\\
\end{array}
\end{equation}

Here I imply no summation over $a$. Also, here $\epsilon_{ac}$ is the antisymmetric tensor.
The extra fields $B_{A}$ and $\lambda_{A}$ are introduced and we have extra transformation rules

\begin{equation}
\begin{array}{ll}
 \delta_{a}c_{Ab}=\epsilon_{ab}B_{A}\\
 \delta_{a}B_{A}=0\\
 \delta_{a}\phi_{A}^{*b}=-\delta_{a}^{b}[(-1)^{a}\lambda_{A}+\\
  +\frac{1}{2}(B_{A}+\frac{\delta^{l}R_{A1}(\phi_{A}-\varphi_{A1}-\varphi_{A2})}{\delta\phi_{B}}R_{B2}(\phi_{B}-\varphi_{B1}-\varphi_{B2}))]\\
 \delta_{a}\lambda_{A}=0\\
\end{array}
\end{equation}

The gauge fixing procedure must occur in a BRST-anti-BRST invariant way.
The inclusion of the terms involving $\frac{\delta^{l}R_{A1}}{\delta\phi_{B}}R_{B2}$ as well as the additional terms in eq. (56)
as a modification of the traditional BRST transformation rules 
is done in order to encode the nilpotency of the BRST-anti-BRST transformation in a way that is
independent of the gauge algebra.
 The additional fields $B_{A}$ and $\lambda_{A}$ have the role of imposing the nilpotency at the level of 
the transformation rules of the original fields. Any off-shell non-nilpotency is thus shifted to the
transformation rules of the collective fields.
There are various ways in which more than one gauge symmetry can be encoded in the BRST transformation rules.
 Also parts of some transformation rules can be transfered to transformation rules of additional fields.
These properties have many possible applications. 
Here I make use of them in order to avoid the sign problem.
 One can introduce a matrix $M^{AB}$ which is invertible and has the property
\begin{equation}
 M^{AB}=(-1)^{\epsilon_{A}\epsilon_{B}}M_{BA}
\end{equation}
It also makes all the entries between the Grassmann odd and Grassmann even sectors vanish.
This means that the term $\phi_{A}M^{AB}\phi_{B}$ has ghostnumber zero and even Grassmann parity.
One can gauge fix to zero the collective terms
\begin{equation}
\begin{array}{ll}
 S_{col}=-\varphi_{A+} M^{AB}\lambda_{B}+\frac{1}{2}\varphi_{A-}M^{AB}B_{B}+\\
 +(-1)^{a}(-1)^{\epsilon_{B}}\phi_{A}^{*a}M^{AB}c_{Ba}+\\
+\frac{1}{2}\varphi_{A-}M^{AB}\frac{\delta^{l}R_{B1}(\phi_{B}-\varphi_{B+})}{\delta\phi_{C}}R_{C2}(\phi_{C}-\varphi_{C+})\\
+(-1)^{a+1}(-1)^{\epsilon_{B}}\phi_{A}^{*a}M^{AB}R_{Ba}(\phi_{B}-\varphi_{B+})
\end{array}
\end{equation}
Here the summation over a is implied. The sum of the two collective fields is fixed to zero, $\phi_{A}^{*a}$ are the source terms for the BRST-anti-BRST transformations
 and the difference between the two collective fields
$\varphi_{A-}$ is the source of the mixed transformations. 
The original gauge symmetry can be fixed in an extended BRST-invariant way by adding the variation of a gauge boson $\Psi(\phi)$ of ghostnumber zero.
\begin{equation}
S_{\Psi}=\frac{1}{2}\epsilon^{ab}\delta_{a}\delta_{b}\Psi(\phi_{A})
\end{equation}
The gauge fixed action can be written as:
\begin{equation}
 S_{gf}=S_{0}[\phi_{A}-\varphi_{A+}]+S_{col}+S_{\Psi}
\end{equation}
where
\begin{equation}
 S_{col}=-\frac{1}{4}\epsilon^{ab}\delta_{a}\delta_{b}(\varphi_{A1}M^{AB}\varphi_{B1}-\varphi_{A2}M^{AB}\varphi_{B2})
\end{equation}
At this moment we have a gauge fixed action with a BRST-anti-BRST symmetry.
Although the matrix $M$ can be eliminated in the end, it has the potential to
introduce a metric on the space spanned by the fields of the trivial system and the ghosts [29].
The same idea can be used to introduce a Kahler structure on the field space (see section 4 in [20] for a review of Kahler structures).
The introduction of the internal space is required due to the definition of the Hodge dual operation. As can be seen in [20] (section 2)
and in [21] the internal space allows the definition of the Hodge operator in any dimension. This will extend the applicability of the procedure 
described in [24] (see also [20], section 3) for arbitrary dimensions of the original theory (considering the suitable generalization of the indices of the operators).
The discrete symmetry is obtained by introducing new collective fields and imposing a dual-space gauge fixing that generates a Kahler structure.
The Hodge star operator that relates the direct and dual emerging global continuous symmetry transformations will induce a discrete symmetry in the final theory. 
This symmetry can be associated to a form of artificial discrete invariance (see example at the end of section 3 in ref [20] or ref [22]-[23]). 
Now I am focusing on the method that generates a time-reversal type symmetry.
One spans an internal space by the introduction of a new set of fields and an equivalent of the $M$ matrix. Take the action
$S_{col}$ used above.
Now introduce an internal space index for the collective fields $\varphi_{Aa}^{\Omega}$, define their dual with respect to the internal space:
\begin{equation}
 \tilde{\varphi}_{Aa}^\Omega=\frac{1}{2}\epsilon_{\Omega\Gamma}\varphi_{Aa}^{\Gamma}
\end{equation}
and rewrite the fields as
\begin{equation}
 \varphi_{Aa}^{\pm\Omega}=\frac{1}{2}(\varphi_{Aa}^{\Omega}\pm i \tilde{\varphi}_{Aa}^{\Omega})
\end{equation}
In the same way introduce another matrix $N$ that can be factorized as:
\begin{equation}
\begin{array}{cc}
N^{\Omega\Gamma}=\frac{1}{2}(h^{\Omega\Gamma}-if^{\Omega\Gamma}) \\
\overline{N}^{\Omega\Gamma}=\frac{1}{2}(h^{\Omega\Gamma}+if^{\Omega\Gamma})\\
\end{array}
\end{equation}
The matrices $f$ and $h$ are completely arbitrary as long as the matrix $N$ can be decomposed in the above way 
(see section 5 and 6 from [20] for a discussion about the role of $N$).
Replacing this into the action together with a corresponding change in the fields we obtain a Kahler structure imposed over the manifold of the field-antifield formalism.
This procedure may be related to the idea of Kahler polarization in the geometric quantization. 
The additional terms obtained in the matrix are now of the form: 
\begin{equation}
 S_{col}=-\frac{1}{4}\epsilon^{ab}\delta_{a}\delta_{b}\delta\bar{\delta}(\varphi_{A1}^{-\Omega}N_{\Omega\Gamma}\varphi_{B1}^{-\Gamma}-\varphi_{A2}^{+\Omega}\overline{N}_{\Omega\Gamma}\varphi_{B2}^{+\Gamma}) 
\end{equation}
where now, the $\delta$ and $\bar{\delta}$ operators correspond to the dual BRST transformations.
Their form depends on the practical calculation. 
The most general expression that can be written here is
\begin{equation}
\begin{array}{l}
 \delta_{Da}\phi_{A}=\phi_{Aa}^{*}\\
\delta_{Da}\varphi_{Ab}=\delta_{ab}[\phi_{Aa}^{*}-\epsilon_{ac}c_{A}^{c}-R_{Aa}]+(1-\delta_{ab})\epsilon_{ac}c_{A}^{c}\\
\delta_{Da}c_{Ab}=-\delta_{a}^{b}[(-1)^{a}\lambda_{A}+\frac{1}{2}(B_{A}+\frac{\delta^{l}R_{A1}}{\delta\phi_{B}}R_{B2})]\\
\delta_{Da}B_{A}=0\\
\delta_{Da}\phi_{A}^{*b}=\epsilon_{ab}B_{A}^{a}\\
\delta_{Da}\lambda_{A}=0\\
\end{array}
\end{equation}
Where the same convention remains valid as for the BRST and anti-BRST transformations.
However, several expressions may be altered according to the particularities of each theory.
The expressions for the case of $2D$ $QED$ are given in section 3 of [20]. $S_{col}$ is the equivalent of the collective term in the action for
the new degrees of freedom constructed to introduce the Kahler structure. The matrices $N_{\Omega\Gamma}$ and $\overline{N}_{\Omega\Gamma}$ allow me to write the gauge fixing term
in such a way that a Kahler structure becomes visible. They may be compared to the choice of a polarization set over the field space although here the scope is another. 
$M^{AB}$ is considered implictly. The same method that allows the $M$ matrix to vanish eliminates the $N$ matrix as well if this is our intention. This would lead to losing the
``polarization'' that makes the Kahler structure visible (please note that I use the term ``polarization'' in a non-rigorous sense refering only to the way in which the fields
can be partitioned). We now have a
 Kahler structure imposed on our original action. The dual-BRST symmetry is the BRST symmetry created by the new collective fields
 together with their trivial system. It is the analogue of the co-derivative from algebraic geometry. In this way we obtained the
 so-called de-Rham cohomology operators that are now identified with the (anti)-BRST and dual-(anti)-BRST operators. 

The connection between them is given by the Hodge star operator which can be constructed independent of the dimension of the original field space if one
 follows the prescription of constructing the internal spaces as described above. In this case the Hodge duality plays the role of a discrete symmetry transformation
 (section 3, ref. [20]).

\par As noted in reference [13] and [15]  the field-anti-field setup is amenable to the construction of a Kahlerian structure imposed on the system of fields. 
 Here, the Hodge star induces a symmetry that can be 
identified with time-reversal in the case of Kahlerian structures.
If one thinks at the antipode in a Hopf algebra one can see that there are not few similarities between the Hodge star operator and the antipode.
All one has to do is to suitably introduce fields and antifields 
via appropriate trivial symmetries such that the antipodal structure (associated to the Hodge dual) becomes visible. 
In the context of the field-antifield approach the structure of the emerging fermionic determinant will be
\begin{equation}
 det(D) = \left( \begin{array}{cc} iT_{1} & 0 \\ 0 & -iT_{1} \end{array} \right)
\end{equation}
where $T_{1}$ results from the construction of the Kahler structure. This assures that for the extended field space the sign of the fermionic determinant is always positive and allows us to avoid the sign problem.
The fact that the imposed structure is reflected on the form of the determinant is explained in section 7 of [20] as well as in ref. [17].
\par In order to asses the complexity of the final problem, considering the fact that the fermionic determinant does not change sign (this can be interpreted in the formalism of the first chapter as the weights $p(c)$ being positive) the increase in complexity is due to the addition of more fields. In the above constructions the number of fields has been doubled two times so I went from a theory containing $N$ fields to a theory containing $4N$ fields. Also, additional fields have been added each time in order to insure the desired gauge fixing. The additional fields on the BRST and anti-BRST branches are related so the construction of the BRST-anti-BRST structure required $4N$ fields and the dual counterpart required another $4N$ fields. 
This amounts to a theory containing $8N$ fields globally. Considering that half of the fields live in the internal space and have a controlled behavior and also that the 
increase in the field number is polynomial,
the method should not add exponential complexity. 
\section{Conclusion}
In conclusion, I showed that it is possible to introduce an auxiliary discrete symmetry that mimics time-reversal and that this symmetry can
 be used in order to avoid the sign problem.
\par The method used here is fairly general and I foresee various applications in computer science, condensed matter, exotic states of matter, etc.
 In general it opens completely new perspectives on notions like symmetry and it identifies a new and interesting connection between geometry (symmetry) and topology. 
 \section {Acknowledgement}
 This work is supported by ERC Advanced Investigator Project 267219.
\\

\title{Supplemental material}
\begin{abstract}
I present in this supplemental material several mathematical constructions required for the proper understanding of the main article. Also, a comparison between the method presented in the main article and a well established alternative method is made. The two methods agree numerically. Although the considered situation is simple and does not present a clear numerical proof for the proposed method, it shows that in the domain where both methods are valid, their results agree. Extensions to regions where the classical method is unavailable are hence desirable. 
\end{abstract}
\maketitle
\section{ Hodge star and Hodge duality}
Let $(M,g)$ be a $N=2d$-dimensional manifold for which we can define the * operator in the following way [16]:
\begin{equation}
\begin{array}{ll}
 \alpha \wedge * \beta =g_{p}(\alpha,\beta)dv_{g}; & \alpha,\beta \in \wedge^{N}\\
\end{array}
\end{equation}
We have also that $(**)=1$ on $\wedge^{N}$ which means that $\wedge^{N}$ splits into eigenspaces as 
\begin{equation}
 \wedge^{N}=\wedge_{+}+\wedge_{-}
\end{equation}
where the two eigenspaces correspond to eigenvalues +1 and -1 respectively. A $d$-form which belongs to $\wedge_{+}$ is called self-dual whereas if it belongs to the other eigenspace it is called 
anti-self-dual. An important remark to be done here is that given a p-vector $\lambda\in\wedge^{p}V$ then $\forall \theta\in\wedge^{d-p}V$ there exists the wedge product such
 that $\lambda\wedge\theta\in\wedge^{d}$.
The (anti)BRST and dual-(anti)BRST operators are then equivalent to the operators:
\begin{equation}
 \delta_{a},\delta_{b}:\wedge^{k}\rightarrow \wedge^{k+1}
\end{equation}
\begin{equation}
 \delta,\bar{\delta}=*\delta_{a,b}*:\wedge^{k}\rightarrow \wedge^{k-1}
\end{equation}
\begin{equation}
 \Delta=\delta_{a,b}(*\delta_{a,b}*)+(*\delta_{a,b}*) \delta_{a,b}:\wedge^{k}\rightarrow \wedge^{k}
\end{equation}
In the context of algebraic geometry these are in order: the exterior differential, the coexterior(dual) differential and the Laplace operator. 
The exact and co-exact forms are orthogonal. 
The Hodge theorem allows the identification of a unique representative for each cohomology class as belonging to the Kernel of the Laplacian defined for 
the specific complex manifold. 
If this is put together with the definition of the Kahler manifold we obtain extra (discrete) symmetries in the Hodge structure of the manifold. 
In the main paper the dual operators acting on the field space have been introduced in a general context. 
For a practical description in the context of field-spaces see ref. [22].
There the author starts from a field theory with physical terms and identifies the discrete symmetry as the one induced by the Hodge-* operator in the physical
 context. In the current approach, the BV formalism generates the usual even dimensional symplectic space. 
Dualization of the BRST-anti-BRST operators in this work is done using the extended symplectic field structure. One is not supposed to assume physicality of the terms involved.

\section{ Internal spaces and duality}
The use of internal spaces in order to naturally define duality operations is not new. In fact I follow here reference [21] to show that the construction of an internal space 
is useful in this context and that a discrete $Z_{2}$ symmetry can appear.

I start by following reference [21] with an example of even dimensional $(2n)$ 
electrodynamics. Let $A$ be a general $(n-1)$ form and $F_{k_{1},...k_{n}}$ its associeated field
strength:
\begin{equation}
 F_{k_{1}...k_{n}}=\partial_{ [ k_{n}}A_{k_{1}...k_{n-1} ] }
\end{equation}
\begin{equation}
 *F^{k_{1}...k_{n}}=\frac{1}{n!}\epsilon^{k_{1}...k_{2n}}F_{k_{n+1}...k_{2n}}
\end{equation}
Given the action, the equation of motion and the Bianchi identity as
\begin{equation}
 S=-c_{n}\int d^{2n}x F_{k_{1}...k_{n}}F^{k_{1}...k_{n}}
\end{equation}
\begin{equation}
 \partial_{k_{1}}F^{k_{1}...k_{n}}=0
\end{equation}
\begin{equation}
 \partial_{k_{1}}*F^{k_{1}...k_{n}}=0
\end{equation}
($c_{n}$ is a constant, $k_{j}$ is the tensorial index) we can see that at the level of the Bianchi identity and the equation of motion the dual operation is a symmetry.
Nevertheless, in general the second power of the dual operation has a different structure depending on the dimension of the space:
\begin{equation}
**F=\left\{ \begin{array}{rcl} F & if & D=4k-2\\-F & if & D=4k \end{array}\right.
\end{equation}
As one can see the dual * is not well defined for the 2-dimensional $(2D)$ scalar or for the 4k-2 dimensional extensions. Its definition has been enlarged [21]
by making an internal structure of the potentials in the theory manifest. 
One should note that this has been achieved by using a canonical transformation and that the same can be achieved via BRST. I will enlarge the set of fields 
(alternatively the Hilbert space) by
giving them an internal structure of the form $(\alpha,\beta)$.
The dual operation is now defined as
\begin{equation}
\begin{array}{ll}
 \tilde{F}^{\alpha}=\epsilon^{\alpha\beta}*F^{\beta},& D=4k
\end{array}
\end{equation}
\begin{equation}
\begin{array}{ll}
\tilde{F}^{\alpha}=\sigma_{1}^{\alpha\beta}*F^{\beta},& D=4k-2
\end{array}
\end{equation}
\begin{equation}
 \tilde{\tilde{F}}=F
\end{equation}
$\sigma_{1}^{\alpha\beta}$ being the first Pauli matrix. In this case self and anti self dualities are well defined in any $D=2k$ dimensional space. 
One can start with the first order form of the theory:
\begin{equation}
 S=\int d^{D}x[\Pi\cdot\dot{A}-\frac{1}{2}\Pi\cdot\Pi-\frac{1}{2}B\cdot B + A_{0}(\partial\cdot\Pi)]
\end{equation}
Maxwell's Gauss constraint can be generalized to be precisely the extended curl $(\epsilon\partial)=\epsilon_{k_{1}k_{2}...k_{D-1}}\partial_{k_{D-1}}$. Then
\begin{equation}
 \Pi=(\epsilon\partial)\cdot\phi
\end{equation}
\begin{equation}
 B=(\epsilon\partial)\cdot A
\end{equation}
where $\phi$ is a $(\frac{d}{2}-1)$-form potential, $A$ is a generalization of the vector potential, $A_{0}$ is the general multiplier that enforces the Gauss
 constraint, the antisymmetrization of $\partial$ is defined as
\begin{equation}
 (\epsilon\partial)=\epsilon_{k_{1}k_{2}...k_{D-1}}\partial_{k_{D-1}}
\end{equation}
and in general the notation 
\begin{equation}
  \Phi\cdot\Psi=\Phi_{[k_{1}...k_{D-1}]}\Psi_{[k_{1}...k_{D-1}]}
\end{equation}
is used to imply antisymmetrization via the brackets.
Now I construct an internal space of potentials where duality symmetry is manifest ($\Phi^{+}$ and $\Phi^{-}$ represent the new field structure). The dual 
projection can be defined now as a canonical transformation of the fields in the following way:
\begin{equation}
 A=(\Phi^{+}+\Phi^{-})
\end{equation}
\begin{equation}
\Pi=\eta(\epsilon\partial)(\Phi^{(+)}-\Phi^{(-)})
\end{equation}
\begin{equation}
 \eta=\pm1
\end{equation}
The action can be rewritten in terms of these fields as
$$
 S=\int d^{D}x\{\eta[\dot{\Phi}^{(\alpha)}\sigma_{3}^{\alpha\beta}B^{(\beta)}+\dot{\Phi}^{(\alpha)}\epsilon^{\alpha\beta}B^{(\beta)}]-B^{(\beta)}\cdot B^{(\beta)}
$$
where $B^{(\beta)}=(\epsilon\partial\cdot\Phi^{(\beta)})$ and $\sigma_{3}^{(\alpha\beta)}$ and $\sigma_{2}^{(\alpha\beta)}=i\epsilon^{(\alpha\beta)}$ are the Pauli marices.
We see that the symplectic part factorizes in two parts: one involving the third Pauli matrix and the other one the second Pauli matrix. 
For a dimension $D=4k$ the first term is the generalization of the $2D$ chiral bosons. The $Z_{2}$ symmetry manifests itself in the 
transformation $\Phi^{(\pm)} \longleftrightarrow \Phi^{(\mp)}$. The second
term becomes a total derivative. For $D=2K$ the first term becomes a total derivative and the second term explicitly shows the symmetry of SO(2).
Although the complete diagonalization of the action in 3D cannot be done in coordinate space a dual projection is possible in the momentum space [21].
Let me introduce a two-basis $\{\hat{e}_{a}(k,x),$ $a=1,2\}$ with $(k,x)$ being conjugate variables and the orthonormalization condition given as
\begin{equation}
 \int dx \hat{e}_{a}(k,x)\hat{e}_{b}(k',x)=\delta_{ab}\delta(k,k')
\end{equation}
The vectors in the basis can be chosen to be eigenvectors of the Laplacian, $\nabla^{2}=\partial \partial$ and
\begin{equation}
 \nabla^{2}\hat{e}_{a}(k,x)=-\omega^{2}(k)\hat{e}_{a}(k,x)
\end{equation}
The action of $\partial$ over the $\hat{e}_{a}(k,x)$ basis is
\begin{equation}
 \partial\hat{e}_{a}(k,x)=\omega(k)M_{ab}\hat{e}_{b}(k,x)
\end{equation}
The two previous equations give
\begin{equation}
 \tilde{M}M=-I
\end{equation}
where $\tilde{M}_{ab}=M_{ba}$.
The canonical scalar and its conjugate momentum have the following expansion
\begin{equation}
 \Phi(x)=\int dk q_{a}(k)\hat{e}_{a}(k,x)
\end{equation}
\begin{equation}
 \Pi(x)=\int dk p_{a}(k)\hat{e}_{a}(k,x)
\end{equation}
where $q_{a}$ and $p_{a}$ are the expansion coefficients. 
The action appears in this representation as a two dimensional oscillator. The phase space is now four dimensional, representing two degrees of freedom per mode,
\begin{equation}
 S=\int dk\{p_{a}\dot{q}_{a}-\frac{1}{2}p_{a}p_{a}-\frac{\omega^{2}}{2}q_{a}q_{a}\}
\end{equation}
now we can introduce the following canonical transformation
\begin{equation}
 p_{a}(k)=\omega(k)\epsilon_{ab}(\varphi_{b}^{(+)}-\varphi_{b}^{(-)})
\end{equation}
\begin{equation}
 q_{a}(k)=(\varphi_{a}^{(+)}+\varphi_{a}^{(-)})
\end{equation}
The action becomes $S=S_{+}+S_{-}$ where
\begin{equation}
 S_{\pm}=\int dk\omega(k)(\pm\dot{q}_{a}\epsilon_{ab}q_{b}-\omega(k)q_{a}q_{a})
\end{equation}
As expected, this action presents the $Z_{2}$ symmetry under the transformation $\varphi_{a}^{\alpha}\rightarrow \sigma_{1}^{\alpha\beta}\varphi_{a}^{\beta}$.

This is a particular example. However, the field-antifield prescription used in the main 
paper has practically a similar role and is defined in general. It generates a symplectic even dimensional field space suitable for quantization.
It also defines an analogues for the
Hodge-* operators.

\section{Hodge star as discrete symmetry}
For an example of how the Hodge star induces a discrete symmetry I follow ref. [22]-[24].
The main idea there was to represent the Hodge decomposition operators $(d,\delta,\Delta)$ as some symmetries of a given BRST invariant Lagrangean of a gauge theory. 
In general, the Hodge decomposition theorem states that on a compact manifold any $n$-form $f_{n}(n=0,1,2,...)$ can be uniquely represented as the sum of a harmonic form $h_{n} 
(\Delta h_{n}=0,dh_{n}=0, \delta h_{n}=0)$, an exact form $de_{n-1}$ and a co-exact form $\delta c_{n+1}$ as
\begin{equation}
 f_{n}=h_{n}+de_{n+1}+\delta c_{n+1}
\end{equation}
where here $d$ is the exterior derivative, $\delta$ is its dual and $\Delta$ is the Laplacian operator $\Delta=d\delta+\delta d$.
In order to identify the dual BRST transformation, one has to observe that while the direct BRST transformations leave the two form $F=dA$ in the construction of a 
gauge theory invariant and transform the Dirac fields like a local gauge transformation, the dual-BRST transformations leave the previous gauge fixing term invariant
 and transform the 
Dirac fields like a chiral transformation.
So, as a practical example, I can start like the authors of [24] from a BRST invariant lagrangean for QED noting that generalizations for non-abelian gauge theories with 
interactions exist in the literature as well. 
\begin{equation}
 L_{B}=-\frac{1}{4}F^{\mu\nu}F_{\mu\nu}+\bar{\psi}(i\gamma^{\mu}\partial_{\mu}-m)\psi-e\bar{\psi}\gamma^{\mu}A_{\mu}\psi+B(\partial A)+\frac{1}{2}B^{2}-i\partial_{\mu}\bar{C}\partial^{\mu}C
\end{equation}
$F^{\mu\nu}$ being the field strength tensor, $B$ is the Nakanishi-Lautrup auxiliary field and $C$, $\bar{C}$ are the anticommuting ghosts.
The BRST transformations that leave this Lagrangian invariant are
\begin{equation}
 \begin{array}{ll}
  \delta_{B}A_{\mu}=\eta\partial_{\mu}C & \delta_{B}\psi=-i\eta e C \psi\\
\delta_{B}C=0 & \delta_{B}\bar{C}=i\eta B\\
\delta_{B}\bar{\psi}=i\eta e C\bar{\psi} & \delta_{B}F_{\mu\nu}=0\\
\delta_{B}(\partial A)=\eta \Box C & \delta_{B}B=0\\
 \end{array}
\end{equation}
where $\eta$ is an anticommuting space-time independent transformation parameter.
Particularizing for the 2 dimensional case the Lagrangian becomes 
\begin{equation}
 L_{B}=-\frac{1}{2}E^{2}+\bar{\psi}(i\gamma^{\mu}\partial_{\mu}-m)\psi-e\bar{\psi}\gamma^{\mu}A_{\mu}\psi+B(\partial A)+\frac{1}{2}B^{2}-i\partial_{\mu}\bar{C}\partial^{\mu}C
\end{equation}
and this can be rewritten after introducing another auxiliary field $\mathcal{B}$ as
\begin{equation}
 L_{\mathcal{B}}=\mathcal{B}E-\frac{1}{2}\mathcal{B}^{2}+\bar{\psi}(i\gamma^{\mu}\partial_{\mu}-m)\psi-e\bar{\psi}\gamma^{\mu}A_{\mu}\psi+B(\partial A)+\frac{1}{2}B^{2}-i\partial_{\mu}\bar{C}\partial^{\mu}C
\end{equation}
The dual BRST symmetry operators to be associated to the theory above in the 2 dimensional case are [24]
\begin{equation}
 \begin{array}{ll}
  \delta_{D}A_{\mu}=-\eta\epsilon_{\mu\nu}\partial_{\nu}\bar{C} & \delta_{D}\psi=-i\eta e \bar{C}\gamma_{5} \psi \\
\delta_{D}C=-i \eta\mathcal{B} & \delta_{D}\bar{C}=0\\
\delta_{D}\bar{\psi}=i\eta e \bar{C}\gamma_{5}\bar{\psi} & \delta_{D}F_{\mu\nu}=\eta\Box\bar{C}\\
\delta_{D}(\partial A)=0 & \delta_{D}B=0\\
\delta_{D}\mathcal{B}=0\\

\end{array}
\end{equation}
Moreover, as noted in reference [24] the interacting Lagrangian in 2 dimensions is invariant under the following transformations
\begin{equation}
 \begin{array}{ll}
C \rightarrow \pm i\gamma_{5}\bar{C} & \bar{C}\rightarrow \pm i \gamma_{5}C\\
\mathcal{B}\rightarrow \mp i\gamma_{5}B & A_{0}\rightarrow\pm i\gamma_{5}A_{1}\\
A_{1}\rightarrow\pm i \gamma_{5}A_{0} & B\rightarrow \mp i \gamma_{5}\mathcal{B}\\
E\rightarrow \pm i \gamma_{5}(\partial A) & (\partial A)\rightarrow \pm i \gamma_{5}E\\
e\rightarrow \mp i e & \psi\rightarrow\psi\\
\bar{\psi}\rightarrow\bar{\psi}\\
\end{array}
\end{equation}
Reference [24] shows that these are the analogues of the Hodge duality $(*)$ for this particular example and that they induce a discrete symmetry. 
One can also verify that
\begin{equation}
 * ( *\Phi )=\pm\Phi
\end{equation}
where for $(+)$ the generic field $\Phi$ is $\psi$, $\bar{\psi}$ and for $(-)$ $\Phi$ represents the rest of the fields. 
One can also observe that for the direct and dual BRST symmetries 
\begin{equation}
 \delta_{D}\Phi = \pm * \delta_{B} * \Phi
\end{equation}
is valid.
It has been known before that the above statements are valid for any even dimensional theory [22] and applications for $D=4$, $(3,1)$ and $D=6$ dimensional theories have been given. 
However, combining the ideas presented in the main paper with the observations in ref. [21] and some theorems of algebraic topology and geometry one can generalize the 
applicability of this method to any dimension. While it is true that in some cases non-local transformations emerge ([25]-[27]) the method described in this paper is simply 
a mathematical trick that allows the construction of dual theories with no sign problems so physical meaning of the artificial transformations is irrelevant. 

\section{ Kahler manifolds}
\par Kahler manifolds are particularly interesting for the current problem. In general 
having a differential manifold $\mathcal{M}$ and a tensor of type $(1,1)$ $J$ such that  $\forall p \in \mathcal{M}$, $J_{p}^{2}=-1$, the tensor $J$ will give a structure
 to $\mathcal{M}$ 
with the property that the eigenvalues of 
it will be of the form $\pm i$.
This means that
 $J_{p}$ is an even dimensional matrix and $\mathcal{M}$ is an even manifold. From the same definition it follows that $J_{p}$ can divide the complexified tangent space 
at $p$ in two disjoint vector subspaces
\begin{equation}
 T_{p}\mathcal{M}^{C}=T_{p}\mathcal{M}^{+}\oplus T_{p}\mathcal{M}^{-}
\end{equation}
\begin{equation}
 T_{p}\mathcal{M}^{\pm}=\{Z\in T_{p}\mathcal{M}^{C}|J_{p}Z=\pm iZ\}
\end{equation}
One can introduce two projection operators of the form
\begin{equation}
 P^{\pm}:T_{p}\mathcal{M}^{C}\rightarrow T_{p}\mathcal{M}^{\pm}
\end{equation}

\begin{equation}
 P^{\pm}=\frac{1}{2}(1\pm iJ_{p})
\end{equation}
which will decompose Z as $Z=Z^{+}+Z^{-}$. This construction will generate a holomorphic and an antiholomorphic 
sector: $Z^{\pm}=P^{\pm}Z\in T_{p}\mathcal{M}^{\pm}$, $T_{p}\mathcal{M}^{+}$ being the holomorphic sector.
A complex manifold appears when demanding that given two intersecting charts $(U_{i},\gamma_{i})$ and $(U_{j},\gamma_{j})$, the map $\psi_{ij}=\gamma_{j}\phi^{-1}_{i}$ from
$\gamma_{i}(U_{i}\cap U_{j})$ to $\gamma_{j}(U_{i}\cap U_{j})$ is holomorphic. 
Here $\gamma_{i}$ and $\gamma_{j}$ are chart homeomorphisms and $\psi_{ij}$ is the transition map.
In this case the complex structure is given independently from the chart by
\begin{equation}
 J_{p} = \left( \begin{array}{cc} 0 & 1 \\ -1 & 0 \end{array} \right) \forall p\in \mathcal{M}
\end{equation}
In the complex case there is a unique chart-independent decomposition in holomorphic and antiholomorphic parts. This means we can now choose as
 a local basis for those subspaces the vector
$(\frac{\delta}{\delta z^{\mu}},\frac{\delta}{\delta\bar{z}^{\mu}})$ where ($z^{\mu},\bar{z}^{\mu}$) are the complex coordinates
such that the complex structure becomes
\begin{equation}
 J_{p} = \left( \begin{array}{cc} i1 & 0 \\ 0 & -i1 \end{array} \right) \forall p\in \mathcal{M}
\end{equation}
If we add a Riemannian metric $g$ to the complex manifold and demand that the metric satisfies $g_{p}(J_{p}X,J_{p}Y)=g_{p}(X,Y), \forall p\in \mathcal{M}$
 and $X,Y\in T_{p}\mathcal{M}$ then the metric is called 
hermitian and $\mathcal{M}$ is called a hermitian manifold. A complex manifold always admits a hermitian metric. 
Using the base vectors of the complexified $T_{p}\mathcal{M}^{C}$ we can always write the metric locally as
\begin{equation}
 g=g_{\mu\bar{\nu}}dz^{\mu}\otimes d\bar{z}^{\nu} +g_{\bar{\mu}\nu}d\bar{z}^{\mu}\otimes dz^{\nu}
\end{equation}
If we have a hermitian manifold $(\mathcal{M},g)$ with $g$ hermitian metric and a fundamental 2-tensor $\Omega$ whose action on vectors $X$ and $Y\in T_{p}\mathcal{M}$ is 
\begin{equation}
 \Omega_{p}(X,Y)=g_{p}(J_{p}X,Y)
\end{equation}
then we call $\Omega_{p}(X,Y)$ a Kahler form. 
With this definition the Kahler form has some very useful properties. 
Firstly it is antisymmetric
\begin{equation}
 \Omega(X,Y)=g(J^{2}X,JY)=-g(X,JY)=-\Omega(Y,X)
\end{equation}
Then it is invariant under the action of the complex structure
\begin{equation}
 \Omega(JX,JY)=\Omega(X,Y)
\end{equation}
and under complexification 
\begin{equation}
 \Omega_{\mu\nu}=i g_{\mu\nu}=0
\end{equation}
\begin{equation}
 \Omega_{\bar{\mu}\bar{\nu}}=i g_{\bar{\mu}\bar{\nu}}=0
\end{equation}
\begin{equation}
 \Omega_{\mu\bar{\nu}}=-\Omega_{\bar{\nu}\mu}=ig_{\mu\bar{\nu}}
\end{equation}
thus leading to
\begin{equation}
 \Omega=ig_{\mu\bar{\nu}}dz^{\mu}\wedge d\bar{z}^{\nu}
\end{equation}
A Kahler manifold is a hermitian manifold $(\mathcal{M},g)$ whose Kahler form $\Omega$ is closed ($d\Omega$=0). $g$ is called a Kahler metric.
The closing condition defines a differential equation for the metric. 
\begin{equation}
 d\Omega = (\delta+\bar{\delta})ig_{\mu\bar{\nu}}dz^{\mu}\wedge d\bar{z}^{\nu}=
\end{equation}
\begin{equation}
\frac{i}{2}(\delta_{\lambda}g_{\mu\bar{\nu}}dz^{\lambda}\wedge dz^{\mu}\wedge d\bar{z}^{\nu})+\frac{i}{2}(\delta_{\bar{\lambda}}g_{\mu\bar{\nu}}-\delta_{\bar{\nu}}g_{\mu\bar{\lambda}})d\bar{z}^{\lambda}\wedge dz^{\mu}\wedge d\bar{z}^{\nu}=0
\end{equation}
This leads to the relations
\begin{equation}
 \frac{\delta g_{\mu\bar{\nu}}}{\delta z^{\lambda}}=\frac{\delta g_{\lambda\bar{\nu}}}{\delta z^{\mu}}
\end{equation}
\begin{equation}
 \frac{\delta g_{\mu\bar{\nu}}}{\delta \bar{z}^{\lambda}}=\frac{\delta g_{\mu\bar{\lambda}}}{\delta \bar{z}^{\nu}}
\end{equation}
The solution of the above equation takes the form 
\begin{equation}
g_{\mu\bar{\nu}}=\delta_{\mu}\delta_{\bar{\nu}}K_{i}
\end{equation}
 on a chart $U_{i}$ included in the manifold $\mathcal{M}$.
$K_{i}$ is called Kahler potential.

\begin{equation}
 K_{i}:U_{i}\rightarrow R
\end{equation}
\begin{equation}
 K_{i}=K_{i}^{*}
\end{equation}
The Kahler form can be locally expressed in terms of the Kahler potential as
\begin{equation}
 \Omega=i\delta\bar{\delta}K_{i}
\end{equation}
The definition given above is the most general one. In the main paper this method will be used for the specific case of the Quantum Monte Carlo phase sign problem.

\section{BRST-anti-BRST, Kahler partitioning and dual gauge fixing}
One important aspect discussed in the main paper is the simultaneous direct and dual gauge fixing of artificial shift symmetries on a complexified space. This is done using some special properties of the matrices $M$ and $N$.
Following reference [7] the matrix $M$ insures the simultaneous gauge fixing of the collective fields in a BRST-anti-BRST invariant way. This matrix must be invertible and may
have complex numbers as entries. While acting on the field space it must have the symmetry property $M^{AB}=(-1)^{\epsilon_{A}\epsilon_{B}}M_{BA}$. It must also insure that 
$\phi_{A}M^{AB}\phi_{B}$ has global ghostnumber zero, where here, $\phi_{A}$ and $\phi_{B}$ are arbitrary fields from the theory. 
In the discussion of reference [7] no other requirements on the $M$ matrix are needed.  
Geometric quantization follows several important steps. The first would be the construction of a symplectic manifold $\mathcal{M}$ of even 
dimension $(dim(\mathcal{M})=2n)$ using the BV, BRST or field-antifield prescriptions. The next step is called ``polarization'' and involves the selection of $n$ directions over this manifold on which the resulting quantum
 states should 
depend. The probably best known polarizations produce the Schrodinger or momentum representations in basic quantum mechanics. These are however not the only ones. 
While the Batalin-Vilkovisky procedure generates the $2n$ dimensional manifold the procedure of generating the $n$ dimensional quantum space has additional freedom. 
This leads to a different form in which the variables (fields) can be partitioned, called the Kahler polarization.
A procedure very similar to the construction of a polarization is used here in order to introduce a complex structure over the symplectic
manifold. This generates a split of the field structure into two distinct components. 
\begin{equation}
\begin{array}{ll}
 T_{(1,0)}=\{v\in T_{x}\mathcal{M}^{\mathbb{C}}|J_{x}(v)=iv\}; &  T_{(0,1)}=\{v\in T_{x}\mathcal{M}^{\mathbb{C}}|J_{x}(v)=-iv\} \\
\end{array}
\end{equation}

One may observe that $M_{AB}$ has the potential to 
induce a specific metric over the field space constructed from the original fields and the additional ghosts, antighosts, ghost-for-ghosts, etc.
In order to use this potential for the current problem I introduce two other matrices
\begin{equation}
\begin{array}{cc}
N^{\Omega\Gamma}=\frac{1}{2}(h^{\Omega\Gamma}-if^{\Omega\Gamma}) \\
\overline{N}^{\Omega\Gamma}=\frac{1}{2}(h^{\Omega\Gamma}+if^{\Omega\Gamma})\\
\end{array}
\end{equation}
Their role is to induce a special gauge fixing that generates a Kahler structure over the field space. That gauge fixing can be done by choosing a metric over the symplectic
BV (or BRST) field space has been shown in reference [29]. Apart from the standard BRST-anti-BRST operators,
algebraic geometry defines also the dual-BRST-anti-BRST operators. These are related to the direct operators via a Hodge star transformation. 
Moreover, the Hodge star operation induces an extra discrete symmetry. The Kahler structure imposed by the $N^{\Omega\Gamma}$ matrices assures that this symmetry is of the
form of an anti-unitary time reversal operation, as required to solve the sign problem (see ref. [6]).
Polarization has two main parts. First it induces a form of partitioning of
the field space in ``momentum'' and ``position'' types variables. Second, it imposes a condition that eliminates half of these variables from the
definition of the wavefunction. In this case the last part is not of interest. For the first part however one can consider the 
manifold $T^{*}\mathcal{M}$ and define a complex basis $\{z_{j},\bar{z}_{j}\}$. The symplectic form becomes $\omega=\frac{1}{2}d\bar{z}_{j}\wedge z_{j}$ and the 
complex structure is defined by the action on the basis as $Jz_{i}=iz_{i}$ and $J\bar{z}_{j}=-i\bar{z}_{j}$. One can chose to partition the
field space according to the complex structure $J$ inducing spaces (blocks) $\mathcal{P}$ spanned by $\{\frac{\delta}{\delta\bar{z}_{j}}\}_{j=1}^{n}$ and 
 anti-spaces $\mathcal{\bar{P}}$ spanned by $\{\frac{\delta}{\delta z_{j}}\}_{j=1}^{n}$. 
This polarization induces exactly a Kahler structure.
A similar idea is used here for partitioning the field space such that the functional determinant becomes partitioned in complex conjugated blocks.

This construction still allows some freedom used in the main article in order to give to the discrete symmetry shown here the form of a time-reversal type symmetry. This 
becomes manifest when one uses the ($N_{\Omega\Gamma},\bar{N}_{\Omega\Gamma})$ matrices in order to induce the Kahler structure over the fields. The next step is simply to 
introduce the fields and the Kahler ``partitioning'' of the fields in the theory as shown in the main article.
\section{Kahler duality transformation and symmetry}
I show here that via a suitable shift in the field space a theory can be constructed that has the precise form as the one given in the main article for the Kahler-extended
formulation.
Let the Lagrangean be 
\begin{equation}
 L=L_{0}+f_{\Omega\Gamma}\varphi^{\Omega}\varphi^{\Gamma}
\end{equation}
The Lagrangean can be extended by shifting terms and fields 
\begin{equation}
\begin{array}{lll}
 \varphi^{\pm \Omega}=\frac{1}{2}(\varphi^{\Omega}\pm i\tilde{\varphi}^{\Omega}), & N^{\Omega\Gamma}=\frac{1}{2}(h^{\Omega\Gamma}-if^{\Omega\Gamma}), & \overline{N}^{\Omega\Gamma}=\frac{1}{2}(h^{\Omega\Gamma}+if^{\Omega\Gamma})\\
\end{array}
\end{equation}
This will extend the $\varphi^{\Omega}$ potential while the matrices $N$ and $\bar{N}$ will mix the extension with the original terms. In this way at a first instance one obtains
\begin{equation}
  L=L_{0}+f_{\Omega\Gamma}\varphi^{\Omega}\varphi^{\Gamma}+h_{\Omega\Gamma}\varphi^{\Omega}\tilde{\varphi}^{\Gamma}
\end{equation}
and in the end
\begin{equation}
L=L_{0}[\Phi]+L_{col}
\end{equation}
where $\Phi$ is the general notation for any field occuring in the theory.
Now one has to gauge fix this by the equation in the main paper 
\begin{equation}
 L_{col}=-\frac{1}{4}\epsilon^{ab}\delta_{a}\delta_{b}\delta\bar{\delta}(\varphi^{+\Omega}N_{\Omega\Gamma}\varphi^{+\Gamma}-\varphi^{-\Omega}\overline{N}_{\Omega\Gamma}\varphi^{-\Gamma}) 
\end{equation}
But again, here one can make use of the freedom in the definition of the matrices $N$ and $\bar{N}$. Combined with the metric induced by the matrix $M$, the dualization and
the Hodge star operator inducing a discrete symmetry one 
can generate a splitting of the field space in blocks such that the final field structure is of the form similar to the Kahler structure.
One observes that it is of no importance what kind of fields one considers (Grassmann or bosonic) because the whole set of original fields is in the end split into two blocks
after the introduction of the Kahler ``partitioning''. As a consequence this method works for theories combining bosons and fermions with no additional problems. In fact due to 
the specific way in which the symplectic and Kahler structures are constructed one can also identify am artificially induced symmetry between fermions and bosons.

\section{The Jacobian}
The two main ideas of this paper (symmetry out of cohomology and dual gauge fixing) define a new way in which symmetry can be regarded. Instead of considering it as
 given by nature, here, some discrete symmetries are used as artificial tools that can be added or removed from the theory. In order to make this clear I use the 
field-antifield formalism. What one usually
 considers when studying theoretical problems are actions that have some of the fields already integrated out. My choice, adapted for the Quantum Monte Carlo 
sign problem is to use the field-antifield approach in an innovative way such that a
 Kahler structure become manifest in the symplectic even dimensional field space. Following this choice a discrete symmetry generated by the Hodge dual $(*)$
 emerges. This symmetry assures that the fermionic
 determinant is positive definite. 
The specific way in which the new structure is induced is by introducing a set of auxiliary fields that can be seen as shifts in the field space. After
 performing two shifts one obtains a BRST-anti-BRST structure constructed in a way that enforces the Schwinger Dyson equations as Ward identities. In
 general the Schwinger Dyson equations are the quantum equations of motion. They are derived as a consequence of the generalization to path integrals of
 the invariance of an integral under a redefinition of the integration variable from $x$ to $x+a$. The BRST-anti-BRST symmetry is used in order to enforce
 precisely this at the level of Ward identities. The dual symmetry is obtained analogously by using an internal space. This method ensures that no divergencies
 in any of the kernel momenta appear.
\par One can also ask if it is possible to perform other initial transformations. The answer is of course yes, but the final symmetry
 must be obtained for the entire structure i.e. the action and the integration measure. Performing the transformation as specified and compensating every time
 for the transformations of the measure will produce the same Kahler structure and the same time-reversal-type symmetry that will be mapped into the resulting functional
determinant [18], [19]. 
\par Let $[dq]$ be my initial measure, $G_{a}$ a transformation of the fields and $S[q]$ be my action.
 $[dq]$ is assumed not to be invariant under $G_{a}$. By construction $S[q]$ is considered invariant and so will also be $S'[q',a]$ where $a$ is the parameter 
of the transformation. One assumes the integration over $a$ as being trivial. Performing the change in variables $q\rightarrow q'$ will affect $[dq]$. The
 resulting transformation will be
\begin{equation}
\int [dq]\rightarrow \int [dq']det\lvert \frac{\partial q_{i}}{\partial q'_{j}}\rvert=\int [dq']det(M_{ij})
\end{equation}
\par Here the measure $[dq']$ is not invariant under the gauge transformations. The determinant of the transformation is also not invariant but the invariance is
 recovered when one combines the two transformations. Then, the gauge fixing procedure can be performed and one obtains the emerging global (anti)BRST symmetry. 
Please note that at this level the Jacobian has no special discrete symmetry. On the dual "branch" one can do the same thing obtaining the dual(anti)BRST symmetry. 
After generating the internal space over which one defines the dual BRST symmetry I introduce the hodge star operation which induces a discrete time reversal type 
symmetry over the entire field space and implicitly over the resulting block-determinant.
\par In order to improve on clarity let's think in the terms of the field-anti-field formalism. For the sake of simplicity the field space can be regarded as a $D$ 
dimensional manifold parametrized by real coordinates $y^{i}=(y^{1},y^{2},...,y^{D})$. After performing the field extension in the sense of Batalin-Vilkovisky the
 space is extended to a $2D$ dimensional manifold of the form $y^{i}=(x^{1},x^{2},...,x^{D},\xi^{1},\xi^{2},...,\xi^{D})$ where $x$ are the bosonic and $\xi$ are 
the fermionic coordinates. This space has a symplectic structure given by a closed non-degenerate 2-form
\begin{equation}
\omega=dy^{j}\wedge dy^{i}\omega_{ij}
\end{equation}
\begin{equation}
d\omega = 0
\end{equation}
Finally an antibracket structure emerges 
\begin{equation}
\{A,B\}=A\partial_{i}^{l}\omega^{ij}\partial_{j}B
\end{equation}
By introducing the internal space in the way explained in chapter 3 of the main article (equations 62-64)  one extends the space again. 
Now $D=2d$ and I define the hodge star operation
 and its associated duality. Having the Kahler structure defined by 
\begin{equation}
J = \left( \begin{array}{cccc} 0 & 1 & 0 & 0\\ -1 & 0 & 0 & 0\\0 & 0 & 0 & 1\\ 0 & 0 & -1 & 0 \end{array}\right)
\end{equation}

and going to a complex coordinate basis 
\begin{equation}
\begin{array}{ccc}
z^{a}=(z^{\alpha},\zeta^{\alpha}) & \bar{z}^{a}=(\bar{z}^{\alpha},\bar{\zeta}^{\alpha}),&\alpha = 1,2,...,d
\end{array}
\end{equation}
\begin{equation}
\begin{array}{cc}
z^{\alpha}=x^{\alpha}+ix^{d+\alpha} & \zeta^{\alpha}=\xi^{\alpha}+i \xi^{d+\alpha}\\
\end{array}
\end{equation}
we obtain a supermanifold with a Kahlerian geometry and an equivalent change in the representation of the antibracket. 
Following reference [17] (for the sake of brevity I will not perform the calculations here again) the change in the metric which amounts
 to the redefinition of the poisson bracket (generalized to the antibracket in our situation)
\begin{equation}
\{f,g\}=\sum_{\alpha\beta}\Omega^{\alpha,\beta}\frac{\partial f}{\partial\eta^{\alpha}}\frac{\partial g}{\partial \eta^{\beta}}
\end{equation}
modifies the expression of the integration measure taking the change of the metric in the definition of the antibracket and mapping it
 onto the structure of the resulting global block-determinant. (see eq. (11)-(15) and (17)-(18) of ref. [17]). This ensures that the
 discrete symmetry affects the resulting determinant in the desired way. 
\par Another way of looking at this discrete symmetry is to consider it as induced by the antipode of a hopf-algebra (the vector space
 analogue of the Hodge star). Only after one constructs the global BRST-anti-BRST and dual-BRST-anti-BRST symmetries will the discrete
 symmetry emerge. The method of constructing the first two symmetries already implies the inclusion of the Jacobian of the considered
 transformations. This will correctly modify action as well as the measure of integration (see [8],[9]).
\par One may also notice that here, I used the de-Rham cohomology and Hodge duality in order to generate a discrete symmetry. Further
 symmetries could be obtained considering other topological properties like cobordism or Morse-surgery. 
\section{Practical calculation}
I present here preliminary results obtained by applying a path integral Monte Carlo method to a simple oscillator-quartic anharmonic potential.
While this is not a numerical proof of validity 
it may be considered as a test for a known case. The results presented in the left figure have been 
obtained using the corrective series expansion in the form of an effective potential [28]. The dual gauge fixing method used to produce the right figure
 had no need for such corrective
expansions and converged to approximately the same values. The number of iterations appear larger in the gauge fixing method but one has to consider that 
in the left figure the cost of constructing the effective corrective potential is not considered. That effective potential calculation introduced several other terms 
and became intractable for higher orders.
\begin{figure}[htb]
 \begin{center}
  \includegraphics[width=180pt]{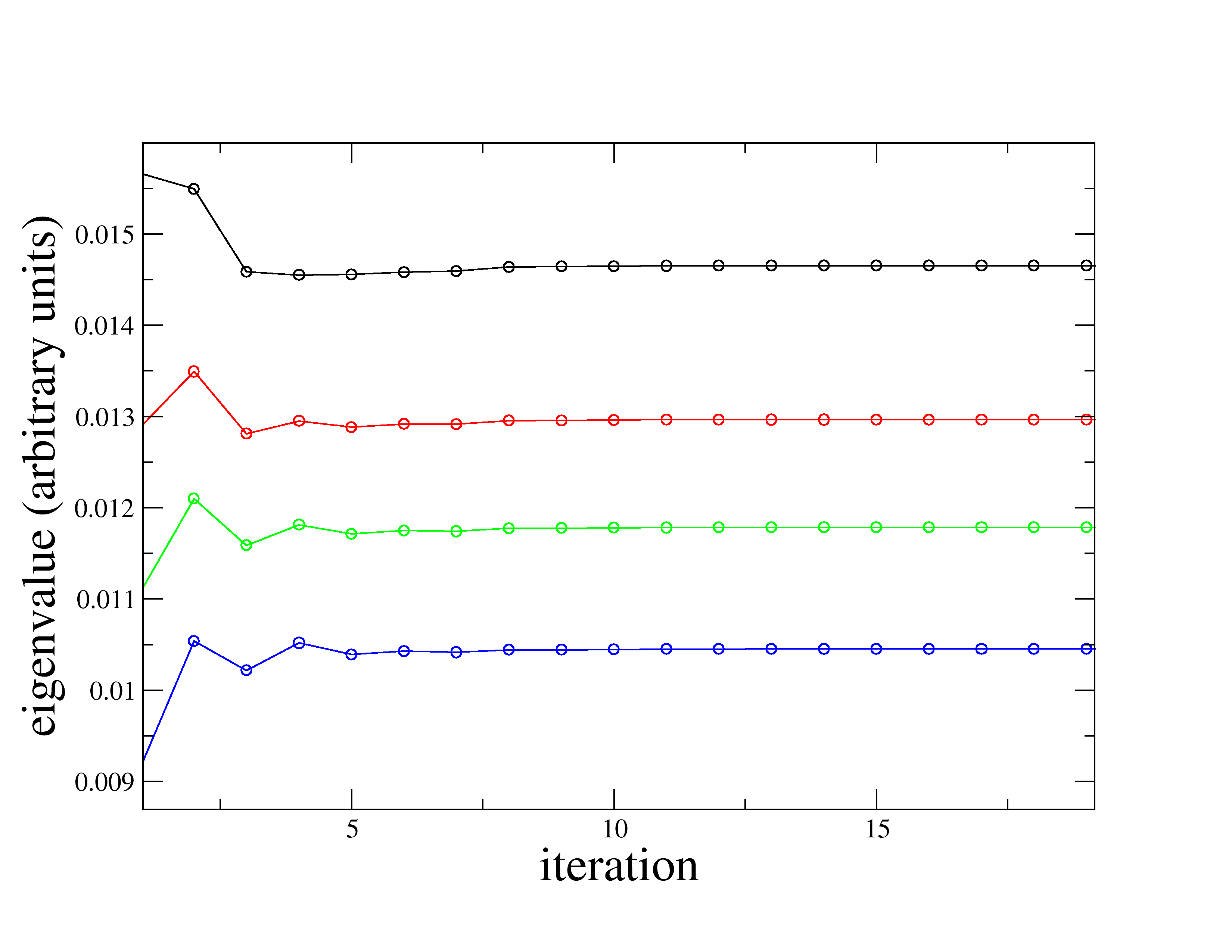}
  \includegraphics[width=180pt]{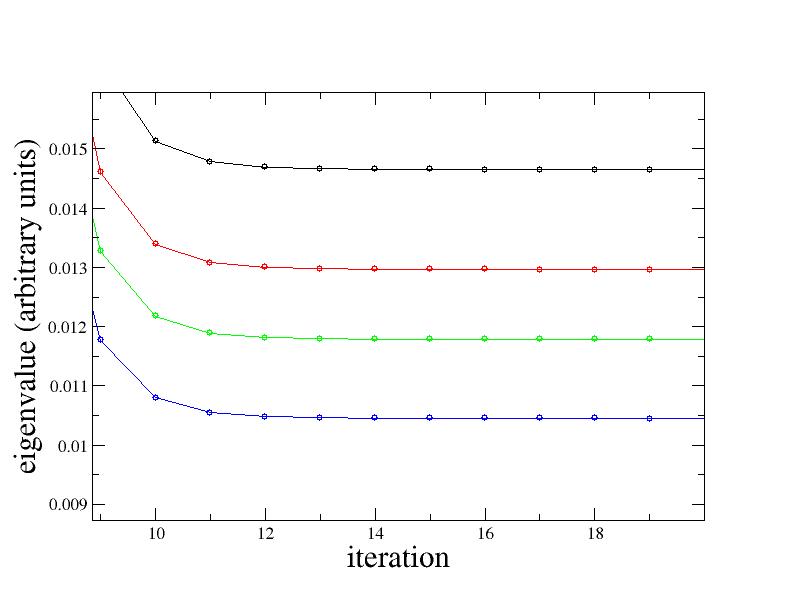}
  \caption{series corrected vs. Hodge symmetric solution}
\end{center}
\end{figure}
I am aware that this example is not specifically related to the fermionic sign problem. The figures presented here aim just to show that my method is valid and consistent
with known results. Further investigation of the effects of this method in more relevant physical situations (especially fermionic problems) is obviously desirable.
\section{bibliography}

\end{document}